\documentclass{egpubl-arXiv}
\usepackage{eg2023}

\ConferencePaper

\usepackage[T1]{fontenc}
\usepackage{dfadobe}  
\usepackage{flushend}
\usepackage{calc}
\usepackage{icomma}

\electronicVersion

\usepackage[pdftex]{graphicx} \pdfcompresslevel=9

\usepackage{xcolor}
\definecolor{classicColor}{HTML}{000000}
\definecolor{metaColor}{HTML}{FF7744}
\definecolor{ourColor}{HTML}{4477FF}

\usepackage{graphicx}
\usepackage{enumitem}
\usepackage{booktabs}
\usepackage[hidelinks]{hyperref}
\usepackage{xspace}
\usepackage{comment}
\usepackage{algorithm}
\usepackage{algpseudocode}

\usepackage{graphicx}
\usepackage{soul}
\usepackage{amsmath}
\usepackage{amssymb}
\usepackage{microtype}
\usepackage{wrapfig}
\usepackage{xcolor}
\usepackage{booktabs}
\usepackage{multirow}
\usepackage{xspace}

\def\figurePath{Figures/}

\def\myfigure#1#2{%
    \begin{figure}[htb]%
    \centering\includegraphics*[width = \linewidth]{\figurePath#1}%
    \vspace{-.2cm}%
    \caption{#2}%
    \label{fig:#1}%
    \end{figure}%
}

\def\mycfigure#1#2{%
    \begin{figure*}[h!]%
    \centering\includegraphics*[width = \linewidth]{\figurePath#1}%
    \vspace{-.2cm}%
    \caption{#2}%
    \label{fig:#1}%
    \end{figure*}%
}

\newcommand{\mywfigure}[3]{%
\begin{wrapfigure}{r}{#2\columnwidth}%
 \vspace{-.2cm}%
  \begin{center}%
    \includegraphics[width=#2\columnwidth]{\figurePath#1}%
    \vspace{-.2cm}%
    \caption{#3}%
    \label{fig:#1}%
    \vspace{-.2cm}%
  \end{center}%
\end{wrapfigure}%
}

\newcommand{\eg}{e.g.,\ }
\newcommand{\ie}{i.e.,\ }
\newcommand{\etal}{et~al.\ }

\newcommand{\refSec}[1]{Sec.~\ref{sec:#1}}
\newcommand{\refFig}[1]{Fig.~\ref{fig:#1}}
\newcommand{\refEq}[1]{Eq.~\ref{eq:#1}}
\newcommand{\refTbl}[1]{Tbl.~\ref{tab:#1}}
\newcommand{\refAlg}[1]{Alg.~\ref{alg:#1}}

\newcommand{\change}[1]{#1}

\soulregister\ref7
\soulregister\cite7
\soulregister\refFig7
\soulregister\refAlg7
\soulregister\cite7
\soulregister\ref7
\soulregister\pageref7
\soulregister\shortcite7
\soulregister\eg0
\soulregister\ie0
\soulregister\etal0

\DeclareGraphicsExtensions{.png,.jpg,.pdf,.ai,.psd}
\newcommand{\mysection}[2]{\section{#1}\label{sec:#2}}
\newcommand{\mysubsection}[2]{\subsection{#1}\label{sec:#2}}

\newcommand{\mymath}[2]{\newcommand{#1}{\TextOrMath{$#2$\xspace}{#2}}}

\usepackage[nolist]{acronym}
\begin{acronym}
\acro{BTF}{Bi-directional Texture Function}
\acro{BRDF}{Bi-directional Reflectance Distribution Function}
\acro{svBRDF}{spatially-varying BRDF}
\acro{CDF}{Cumulative Distribution Function}
\acro{NN}{Neural Network}
\acro{CNN}{Convolutional Neural Network}
\acro{MSE}{Mean-square Error}
\acro{PCA}{Principal Component Analysis}
\acro{SGD}{Stochastic Gradient Descent}
\acro{MLP}{Multi-Layer Perceptron}
\end{acronym}

\newcommand{\R}{\mathbb{R}}

\newcommand{\method}[1]{\texttt{\textbf{#1}}\xspace}
\newcommand{\methodRandom}{\method{Random}}
\newcommand{\methodMeta}{\method{Meta}}
\newcommand{\methodOur}{\method{Ours}}
\newcommand{\methodNielsen}{\method{NJR15}}

\newcommand{\model}[1]{\textsc{\textbf{#1}}\xspace}
\newcommand{\modelPhong}{\model{Phong}}
\newcommand{\modelCookTorrance}{\model{Cook-Torrance}}
\newcommand{\modelCookTorranceShort}{\model{Cook-T.}}
\newcommand{\modelLinear}{\model{Linear}}
\newcommand{\modelNeural}{\model{Neural}}

\newcommand{\myparagraph}[1]{\noindent\textbf{#1}\ }

\newcommand{\scientific}[2]{$#1\!\times\!10^{#2}$}

\newcommand{\defaultOrders}[0]{five orders\xspace}

\title{Learning to Learn and Sample BRDFs}

\author[Liu, Fischer, Ritschel]{
 Chen Liu\qquad 	
 Michael Fischer\qquad 
 Tobias Ritschel
 \\
 \hspace{-0.9cm} University College London      
}

\teaser{
  \centering
  \includegraphics[width=\textwidth]{Figures/Teaser}%
  \caption{Equal-time comparison between a neural BRDF model fitted without meta-learning \textbf{(top)}, fitted with meta-learning \textbf{(second row)} and fitted with our meta-sampling \textbf{(third row)}, all at 64 BRDF acquisition samples.
  Meta-sampling improves the visual quality, as seen right: for the same compute and acquisition time at deployment, the third row is closer to the reference in the fourth row than the second row.}
  \label{fig:Teaser}
}%

\MakeRobust{\Call}

\begin{document}
\maketitle

\begin{abstract}
We propose a method to accelerate the joint process of physically acquiring and learning neural \ac{BRDF} models.
While \ac{BRDF} learning alone can be accelerated by meta-learning, acquisition remains slow as it relies on a mechanical process.
We show that meta-learning can be extended to optimize the physical sampling pattern, too.
After our method has been meta-trained for a set of fully-sampled \acp{BRDF}, it is able to quickly train on new \acp{BRDF} with up to \defaultOrders of magnitude fewer physical acquisition samples at similar quality.
Our approach also extends to other linear and non-linear \ac{BRDF} models, which we show in an extensive evaluation.
\end{abstract}

\mysection{Introduction}{Introduction}
Learned representations of \acp{BRDF} \cite{nicodemus1992geometrical} offer intuitive editing, compact storage or interpolation of material appearance \cite{rainer2019neural,hu2020deepbrdf,rainer2020unified,sztrajman2021neural,fan2021neural}.
What neural \ac{BRDF} models so far do not offer is a way to accelerate acquisition.
Acquisition is slow, because it is a physical process, where a device has to change the illumination and capture an optical response, involving mechanical effort.
Therefore, the simplest way to decrease capture time is to take fewer acquisition samples.
Reducing the number of samples was \change{investigated} for linear \ac{BRDF} models \cite{nielsen2015optimal,ngan2005experimental}.

In this work, we reduce the number of \ac{BRDF} acquisition samples by jointly learning the sample pattern and a non-linear, deep, neural-network based \ac{BRDF} model.
We use meta-learning \cite{finn2017model} to optimize the hyper-parameters of an optimization.
Furthermore, we extend the Metappearance approach \cite{fischer2022metappearance} to also meta-optimize over the sample pattern (``meta-sampling''), reducing the sample count by \defaultOrders of magnitude at similar visual quality. 
Finally, we show that our idea is applicable to \acp{NN} as well as to classic models, such as Phong or mixtures of basis \acp{BRDF}.
\change{Our code is available at \url{https://github.com/ryushinn/meta-sampling}.}

\mysection{Previous Work}{PreviousWork}

\myparagraph{BRDFs}
Classic BRDF models include Phong \cite{phong1975illumination}, Cook-Torrance \cite{cook1982reflectance}, Ward \cite{lafortune1997non} or Disney's shading model \cite{burley2012physically}.
These models are compact to store, \change{lend themselves well to manipulation, but face limitations when it comes to reproducing captured materials}.

\myparagraph{Acquisition}
Gonioreflectometers measure the reflectance, depending on the light and view direction
\cite{erb1980computer,white1998reflectometer,li2006automated,mcallister2002generalized} but this process remains slow, as it requires the mechanical change of light and sensor position.
For spherical objects, this process can be accelerated by imaging all normals at the same time \cite{marschner1999image}.
We consider acquisition a black box that requires effort (time, energy, heat, etc) linear in the number of acquisition samples.
Our aim is to reduce this effort.

BRDF acquisition setups have led to the construction of BRDF databases
\cite{ngan2005experimental,matusik2003data}, which we will rely on in this work.

Fitting parametric BRDF models requires an optimization \cite{lensch2003image,nam2018practical}, often with complex target functions, a (differentiable) image formation model and many resulting non-linearities.
We operate on another layer of abstraction, and ask how to automatically tune this optimization on some training BRDFs, together with the BRDF acquisition's sampling so they jointly perform best on new test \ac{BRDF} optimizations, \ie on unseen tasks. 

\myparagraph{Optimizing acquisition}
Several approaches have sought to reduce the BRDF capture time, for example using adaptive sampling
\change{\cite{fuchs2007adaptive,dupuy2018adaptive}}.
Most related to our work is the linear statistical analysis of a set of BRDFs \cite{nielsen2015optimal}.
Authors optimize for a sample pattern, assuming the BRDF they wish to reconstruct can be expressed as a linear combination of basis BRDFs, found using \ac{PCA}.
\change{\cite{lawrence2004efficient} and \cite{dupuy2018adaptive} derive BRDF models that also lead  more efficient acquisition.}
\change{Acquisition can be accelerated further when using a more principled objective function \cite{bieron2020adaptive}.}
Similar ideas were proposed for \acp{svBRDF}
\cite{zhou2016sparse,yu2016sparse} and for \acp{BTF}
\cite{den2018rapid}.
Our work differs in that it targets non-linear, deep, representation of \acp{BRDF}.

For \acp{svBRDF}, light patterns have been optimized together with an auto-encoder for reconstruction \cite{kang2019learning,kang2018efficient}.
Similar ideas apply to image-based relighting \cite{xu2018deep}, related to BRDFs.

\myparagraph{Deep BRDF representation}
Recently, methods have been proposed to represent the BRDF itself by a neural network
\cite{rainer2019neural,hu2020deepbrdf,rainer2020unified,sztrajman2021neural,fan2021neural}.
These methods can be more expressive and offer improved editing or interpolation properties.
However, fitting them to a new BRDF can be time-consuming for two reasons: first, a lengthy optimization is required, and second, the fitting process makes use of many BRDF samples.
For example, \cite{sztrajman2021neural} use over \scientific{8}{5} samples to learn a \ac{BRDF} instance and \cite{hu2020deepbrdf} even use 100\% of the measurements in a BRDF. 
In our work we combine the idea of optimizing the optimization with also optimizing over the sampling.

\myparagraph{Deep material acquisition}
A popular approach to speed up acquisition of (sv)BRDFs is to learn a mapping from images to BRDFs, either supervised
\cite{rematas2016deep,georgoulis2017reflectance,deschaintre2020guided,deschaintre2018single,liu2017material}, or with some level of self-supervision and differentiable rendering in the mix
\cite{park2020seeing,gao2019deep,guo2020materialgan,henzler2021neuralmaterial}.
These methods produce parameters to classic BRDF models (and inherit their limitations), while we produce a \ac{NN} that represents the BRDF itself.

\myparagraph{Meta-learning}
\change{From the previous paragraphs}, we see that two schools exist on how to represent and acquire BRDFs: either by learning feed-forward networks, typically a CNN that maps images to parameters, or by running optimizations on many carefully calibrated measurements.
The first is fast but with limited quality, the second takes longer, but provides better quality.
One proposal to bridge this gap is meta-learning.
It uses an optimization at test-time to fit to observations, but this optimization has been optimized on a training set of many optimization tasks \cite{finn2017model}.
This idea has been applied in computer vision
\cite{sitzmann2020metasdf,wang2021metaavatar,bergman2021fast,tancik2021learned} and also to visual appearance \cite{maximov2019deep,fischer2022metappearance}.
These methods assume the training data given and then learn how to optimize on it.
In this work we take this further, and also optimize over what the training data needs to be.
So instead of learning to optimize with a given set of BRDF samples, we also learn what BRDF samples to take to then fit successfully.


\myparagraph{Sampling strategies}
Learning to sample is not novel in the deep learning community and research on the order of feeding samples can date back to the framework of Curriculum Learning \cite{bouchard_online_2016}.
\cite{stich_safe_2017, zhao_stochastic_2015, needell_stochastic_2014, katharopoulos_not_2019} wish to learn an ideal probability distribution, from which \ac{SGD} draws training samples in terms of importance sampling, so the stochastic gradient with limited samples will benefit from the samples' reduced variance. 
They are expected to obtain the (sub)optimal approximation of the unbiased gradients given a set of samples, but there is no guarantee that the unbiased first-order gradient would guide the best path of optimization.

Active or online sampling research \cite{sener_active_2018,diolatzis2022active} chooses to generate or label the next sample based on metrics inspired by Curriculum Learning, \eg gradient norm or loss, to allocate more training time to the samples that will reduce the loss the most.
The sampling pattern we find transfers across problem instances and does not need to be re-learned.
We operate at a regime of extremely few samples and use orders of magnitudes fewer samples than the aforementioned methods. 
\mysection{Meta-sampling}{OurApproach}

We will first recall classic fitting of a BRDF
(\refSec{Random}), then how meta-learning extends this (\refSec{Meta}) and finally introduce our contribution: joint meta-learning of the BRDF model's parameters and the samples to take (\refSec{Sampling}).

\mymath{\tasks}{\mathcal T}
\mymath{\task}{T}
\mymath{\samples}{\xi}
\mymath{\sample}{\mathbf x}
\mymath{\sampleValue}{\mathbf y}
\mymath{\sampleCount}{n}
\mymath{\optimizer}{\phi}
\mymath{\modelParameters}{\theta}
\mymath{\stepSize}{\Delta}
\mymath{\metaStepSize}{\stepSize_\mathrm o}
\mymath{\sampleStepSize}{\stepSize_\mathrm s}
\mymath{\metaStepCount}{n_\mathrm o}
\mymath{\sampleStepCount}{n_\mathrm s}
\mymath{\learnStepCount}{n_\mathrm l}

\begin{algorithm}[htb]
    \caption{Learning to learn and to sample  BRDFs.
    The function \textsc{Grad} computes the gradient of the first argument w.r.t. the second.}
    \begin{algorithmic}[1]
    \color{metaColor}
    \Procedure{LearnToLearnAndSample}{\tasks}
        \State \optimizer = \Call{Uniform}{\null}
        \State \samples = \Call{Uniform}{\null}
        \For{$i\in[1, \metaStepCount]$} \label{alg1_startOptimLoop}
            \State \task = \Call{SampleTask}{\tasks}
            \State \modelParameters = \Call{Learn}{%
                \optimizer,
                \samples,
                \task
            }
            \State
            \optimizer = 
            $\optimizer - \metaStepSize
            \cdot$
            \Call{Grad}{%
                \Call{Evaluate}{\modelParameters, \task},
                \optimizer
            } \label{alg1_sgdstep0}
        \EndFor  \label{alg1_endOptimLoop}
        \color{ourColor}
        \For{$i\in[1, \sampleStepCount]$} \label{alg1_startSampleLoop}
        \State \task = \Call{SampleTask}{\tasks}
        \State \modelParameters = \Call{Learn}{\optimizer, \samples, \task}
            \State
            \samples = 
            $\samples - \sampleStepSize\cdot$
            \Call{Grad}{%
            \Call{Evaluate}{%
                \modelParameters,
                \task
            }, \samples} \label{alg1_sgdstep1}
        \EndFor  \label{alg1_endSampleLoop}
        \color{metaColor}
        \State\Return \optimizer, \textcolor{ourColor}{\samples}
    \EndProcedure
    \vspace{0.2cm}
    \color{classicColor}
    \Procedure
    {Learn}{\optimizer, \samples, \task}
    \State \modelParameters  = $\optimizer_\mathrm{init}$
    \For{$i\in[1, \learnStepCount]$}
        \State $\sample = \Call{Sample}{\samples}$
        \State $\modelParameters  = 
        \modelParameters  - \optimizer_\mathrm{step} \cdot$ \label{alg1_sgdstep2}
        \Call{Grad}{\Call{Loss}{\sample, \modelParameters , \task}, \modelParameters }
    \EndFor
    \State\Return\modelParameters 
    \EndProcedure  
    
    \vspace{0.2cm}
    
    \Procedure{Evaluate}{\modelParameters, \task}
        \State \Return \Call{Loss}{\Call{Sample}{\Call{Uniform}{\null}}, \modelParameters, \task}
    \EndProcedure

    \vspace{0.2cm}
    \Procedure{Loss}{\sample, \modelParameters , \task}
        \State $c = \cos(\sample.\theta_\mathrm i)$
        \State\Return
        $|
            \log(1+\task(\sample)\cdot c)-
            \log(1+f_r(\sample;\modelParameters)\cdot c
            )
        |$
    \EndProcedure
    \end{algorithmic}
    \label{alg:Method}
\end{algorithm}

\mysubsection{Random}{Random}
We denote classic learning as ``Random'', as the training samples are drawn \change{uniformly} and the model is \change{randomly initialized}.
It uses the function $\textsc{Learn}$ in \refAlg{Method}, which is provided with hyper-parameters \optimizer (e.g., step size, initialization), parameters of a sampling method \samples and a BRDF we want to learn \task\footnote{ 
In the meta-learning literature, learning a single BRDF would be called  a ``task'', hence the symbol.}.
It returns the model parameters \modelParameters  that best encode this BRDF \task.

The function \textsc{Learn} implements learning via stochastic gradient descent:
At each learning iteration $i$, $i \in \{1, 2, ..., \learnStepCount\}$, the sampler, parameterized by \samples, generates samples \sample, which are pairs of incoming and outgoing 3D directions in a suitable parametrization (we use Rusinkiewicz angles \cite{rusinkiewicz1998new}).
The function \textsc{Loss} then queries the BRDF supervisions and compares it with model's predictions for those samples.
We adopt the mean absolute logarithmic error of the  cosine weighted reflectance values proposed in \cite{sztrajman2021neural} as our loss and descend along the loss gradient to update the parameters \modelParameters, usually until convergence (i.e., in classic learning, \learnStepCount is large).
This chain of update steps starts at a certain model initialization (\eg Kaiming-He) and uses a certain learning rate, or step-size, both defined by the hyper-parameter vector \optimizer.

Classic learning usually does not have the capability of automatically determining good parameters for neither \optimizer nor \samples. Usually, hyperparameters like learning rate and model initialization are determined empirically, and samples are drawn randomly from the dataset.  
We will next look into the first issue, while solving the latter is the contribution of this paper.

\mysubsection{Meta}{Meta}
Meta-learning is depicted in \textcolor{metaColor}{orange} in \refAlg{Method}.
Gradient-based meta-learning (we use the MAML algorithm \cite{finn2017model}) relies on a nested optimization, where the inner optimization loop is tasked with over-fitting the model onto a specific \ac{BRDF} \task under the constraint of a very limited number of gradient steps (typically around 10 only, i.e., \learnStepCount is small).
After completion of this inner loop, the model's final performance is evaluated through the function \textsc{Evaluate}, which calculates losses at a set of \change{uniform-}random test samples.
The outer loop then computes the gradient of the performance w.r.t.\ the meta-parameters \optimizer (commonly, learning rate and model initialization), and moves these in a direction that will yield improved performance of the next inner-loop iteration. 
The outer loop subsequently samples a new task \task from a set of tasks \tasks and the inner loop starts anew, thereby ensuring parameters \optimizer that will generalize across all tasks in \tasks. 

We follow the approach of Metappearance \cite{fischer2022metappearance} who meta-learn the initialization and step size for a deep BRDF representation.
Note that the sampling pattern \samples is still random.

\mysubsection{Meta-Learned sampling}{Sampling}
With these steps laid out, we can now define our contribution in \textcolor{ourColor}{blue}.
We repeat the meta-construction from \refSec{Meta}, but now include the parameters of our sampling method \samples in the meta-optimization.
To do so, we use \samples to sample a set of \sampleCount BRDF entries (\ie obtain many supervision pairs of angle pairs \sample and BRDF values \sampleValue) at the start of the inner loop, all from the same BRDF task. 
These are the only samples that the learner will have access to within this inner loop completion, and they are fixed throughout the inner loop. 
Hence, the final model's performance is directly related to the samples produced by \samples, which now allows to compute a meta-gradient w.r.t. \samples. 
As previously done with \optimizer, we use this meta-gradient to adjust \samples such that during the next iteration of the inner loop, the final performance will improve.
This is repeated \sampleStepCount times.
Typically \sampleStepCount is much larger than \sampleCount, \ie the meta-sample-learning will see many tasks (training \acp{BRDF}) for multiple times (epochs).

Experimentally, we found it necessary to decouple the optimization loops of the optimizer's parameters \optimizer and the sampling method \samples to stabilize training and hence run them consecutively (L. \ref{alg1_startOptimLoop}-\ref{alg1_endOptimLoop} and L. \ref{alg1_startSampleLoop}-\ref{alg1_endSampleLoop} in \refAlg{Method}, respectively). 

Note that this is more than simply augmenting \optimizer by some additional dimensions as the samples are not part of the inner-loop optimization (recall, they are sampled once and then fixed), but part of the \textit{supervision} that drives the inner-loop optimization.
So we do not only change the way we learn, given the problem, but we also change the problem (i.e. which samples are chosen from the BRDF) such that it can be learned better.
We hence aim to discover the subset of BRDF samples that is most informative to our learner. 


The sampling method \samples can be parameterized in multiple ways.
During classic- and meta-learning, as already mentioned, \samples is a random generator without any learnable parameters, and \textsc{Sample}(\samples) simply returns random uniform numbers.
For our learned sampling, we went with the most direct approach and parameterized \samples as an explicit \sampleCount-dimensional vector of sample coordinates, which we meta-initialized with a low-discrepancy sequence in 3D.
\samples then is a set, and not a sequence, as batched \ac{SGD} averages over gradients from all samples, which is a symmetric operation that makes order irrelevant.
Meta-learning \samples then becomes as simple as moving the sample coordinates in small steps, so that after each step, \samples becomes slightly more useful to the inner-loop \ac{BRDF} learner.

\mymath{\incoming}{\omega_\mathrm{in}}
\mymath{\outgoing}{\omega_\mathrm{out}}
\mymath{\barrier}{\textsc{Loss}_\mathrm{B}}

\mywfigure{Projection}{0.4}{Projection.}
\myparagraph{Enforcing valid samples.}
Not all vectors \sample in the unit cube $[0,1]^3$ are valid Rusinkiewicz samples, as some configurations result in views below the horizon (the white regions in \refFig{Projection}). 
As a consequence, optimizing \samples might result in individual samples moving into these invalid regions.
Simply redefining the BRDF to have a specific constant value in these regions (e.g., -1.0) leads to areas where the gradient w.r.t. the sample position is zero. This will ultimately result in wasted samples that cannot adjust any longer, as a sample loses all gradient information as soon as it reaches such a constant region and cannot recover. 
Instead of redefining the loss, we opt to extend it by a barrier function \barrier that forces invalid samples back into the defined regions.
One way to achieve this is to penalize their distance to the origin in the $\phi_\mathrm d$-plane (\refFig{Projection}), as in
\begin{align}
    \barrier(\sample, \modelParameters , \task)
    =
    \begin{cases}
    \textsc{Loss}(\sample, \modelParameters , \task) & 
    \text{if } \sample \text{ is valid, and} \\
    \sample.\theta_\mathrm{h}^2 + \sample.\theta_\mathrm{d}^2&
    \text{if else.}
    \end{cases}
\end{align}

This function, applied to all individual samples, is used as a drop-in replacement for \textsc{Loss} in \refAlg{Method}. Also, all the random uniform samples mentioned above and henceforth
are in fact uniform in the valid region, rather than in the cube, which can be achieved by rejecting invalid samples.

\mysubsection{Models}{Models}
To underline the generality of our proposed algorithm, we show that meta-sampling can increase performance on \change{four} increasingly complex BRDF models (\modelPhong, \change{\modelCookTorrance}, \modelLinear and \modelNeural). 
We will now shortly describe these models. 

\mymath{\diffuse}{k_\mathrm d}
\mymath{\specular}{k_\mathrm s}
\mymath{\glossiness}{q}
\mymath{\ksum}{k_\mathrm{sum}}
\mymath{\kratio}{k_\mathrm{ratio}}
\mymath{\lsum}{\lambda_\mathrm{sum}}
\mymath{\lratio}{\lambda_\mathrm{ratio}}

\myparagraph{Phong}
One of the simplest, yet widely used  BRDF models is the physical version \cite{lafortune1994using} of the \modelPhong  \cite{phong1975illumination} model:
\begin{equation}
    f_\mathrm r(\sample)
    =
    \diffuse \frac{1}{\pi} + \specular \frac{\glossiness+2}{2\pi}
    \max(
    \langle 
    \sample.\omega_\mathrm o, 
    \sample.\mathbf r
    \rangle, \,0.0)^\glossiness \,, 
\end{equation}
where $\langle \cdot \rangle$ denotes the dot product between outgoing and reflected direction $\sample.\omega_\mathrm o$ and $\sample.\mathbf r$, respectively.
We re-parametrize \diffuse and \specular by 
\begin{align*}
    &\diffuse = \ksum\kratio\\
    &\specular = \ksum(1-\kratio) \textrm{, where } \\
    &\ksum = \sigma(\lsum) \quad \textrm{ and } \quad \kratio = \sigma(\lratio) \textrm{.} 
\end{align*}
$\sigma$ is the Sigmoid function and hence ensures $\diffuse + \specular \leq 1$. 
Furthermore, we linearize \glossiness by means of an exponential mapping.
The learnable parameters hence are $\lambda_{\textrm{sum}}, \lambda_{\textrm{ratio}} \in \R^3$ and the scalar glossiness \glossiness.
We meta-learn their initial values and a learning rate per parameter.

\mymath{\surfaceNormals}{N}
\mymath{\geometricAttenuation}{G}
\mymath{\fresnelTerm}{F}
\mymath{\fresnelnormal}{F_0}
\mymath{\roughness}{\alpha}
\myparagraph{Cook-Torrance}
\change{The \modelPhong model is easy to implement and cheap to evaluate, but often does not produce realistic appearance, which is why we include the more sophisticated \modelCookTorrance \cite{cook1982reflectance} model in our experiments. \modelCookTorrance explicitly defines the characteristics of a surface's normal distribution \surfaceNormals (we use Beckmann), the geometric attenuation \geometricAttenuation in the surface, and the Fresnel effect \fresnelTerm, for which we use Schlick's approximation \cite{schlick1994inexpensive}. As also implemented in \cite{ngan2005experimental}, we compute the reflectance as 
\begin{equation}
    f_\mathrm r(\sample)
    =
    \diffuse \frac{1}{\pi} + \specular \frac{D(\roughness, \sample)G(\sample)F(\fresnelnormal, \sample)}{\pi\cos(\sample.\theta_\mathrm i)\cos(\sample.\theta_\mathrm o)}.
\end{equation}
The learnable parameters are \diffuse, \specular, roughness \roughness, and the Fresnel value \fresnelnormal. We use a Sigmoid to constrain them within (0, 1).}

\mymath{\basis}{\mathsf A}
\mymath{\basisWeights}{\mathbf w}
\mymath{\basisCount}{m}
\mymath{\reducedBasis}{\hat\basis}
\mymath{\regularizerWeight}{\eta}
\mymath{\mean}{\mu}
\myparagraph{Linear} 
The next higher level of complexity is a \modelLinear model, as proposed by \cite{ngan2005experimental} and refined by \cite{nielsen2015optimal}.
Here, every BRDF is a linear combination of \basisCount basis BRDFs: 
\begin{equation}
    \label{pca_reconstruction}
    f_\mathrm{r}(\sample)
    =
    (\basis \cdot 
    (\basisWeights |1)^\mathsf T)
    [\sample]\,,
\end{equation}
where 
\basis is an (affine) matrix of \basisCount basis BRDFs in Rusinkiewicz parametrization and their mean,
$\basisWeights \in \R^\basisCount$ is a weight vector and
$\left[ \cdot \right]$ is a 3D table lookup.
We follow the \ac{PCA} method from \cite{nielsen2015optimal} and \cite{ngan2005experimental} to construct \basis from MERL, which is the same for all BRDFs.
\basisCount typically is a small number, like 5 we employ here.
The weight vector \basisWeights changes for each BRDF and is fitted to \sampleCount \ac{BRDF} observations $\sampleValue\in\R^\sampleCount$ at \sampleCount direction pairs \sample in closed form:
\begin{equation}
\label{eq:CloseForm}
\basisWeights=
(
\reducedBasis^\mathsf T\reducedBasis+
\regularizerWeight \mathsf  I
)^{-1}
\reducedBasis^\mathsf T \sampleValue,
\end{equation}
where
$\reducedBasis\in\R^{\sampleCount\times\basisCount}=\basis[\sample]$ is the ``reduced'' basis, a lookup into the full basis at \sample.
\regularizerWeight is a regularization weight, set to $\regularizerWeight=40$, as proposed in \cite{nielsen2015optimal}.
The closed-form \refEq{CloseForm} replaces the SGD loop in function $\textsc{Learn}$, so there are no meta-learnable SGD parameters for the \modelLinear model.
However, optimizing the samples still impacts the quality, even if the solution is closed-form.
\regularizerWeight could be meta-tuned, but we did not explore this.
Our approach is general enough to support non-linearity, but also flexible enough to efficiently support a closed-form special case.

\myparagraph{Neural networks} An even more sophisticated way of encoding BRDF data is non-linear \modelNeural networks.
In our case, we want the network the encode the mapping from light- and view-direction to reflectance data.
As in \cite{sztrajman2021neural}, we use a simple two-layer \ac{MLP} with 21 hidden units per layer, ReLU activations, and an exponential activation for the final layer.
We use this architecture as it is simple and efficient (only 675 trainable parameters), yet highly expressive.
A low number of parameters is desirable, as computing higher order gradients in the meta-learning inner loop is memory-intensive. 
The learnable parameters are the model's weights, fitted to each new BRDF.
The meta-learnable parameters are the initialization and step sizes.

\mysubsection{Implementation}{Implementation}

\mymath{\batchSize}{b}
\mymath{\optimizerBatchSize}{\batchSize_\mathrm o}
\mymath{\sampleBatchSize}{\batchSize_\mathrm s}
\mymath{\modelBatchSize}{\batchSize_\mathrm l}

We implement our method in PyTorch \cite{paszke2019pytorch} and use the learn2learn framework \cite{Arnold2020-ss} as our meta-learning library. 
The outer-loop optimization for the meta-learned optimization happens via Adam with learning rate \scientific{1}{-4} (note that the Adam step was replaced by vanilla SGD in \refAlg{Method}, Lines \ref{alg1_sgdstep0} and \ref{alg1_sgdstep1}, for ease of exposition). 
For the optimization of the inner loop, we use MetaSGD \cite{li2017meta}, \ie optimize a per-parameter learning rate together with the respective method's parameters. 
MetaSGD is initialized with \scientific{1}{-3}, and we run $20$ inner-loop steps of MAML optimization. 

To train \samples, we also use Adam in the outer-loop with learning rate \scientific{5}{-4} and a cosine annealing scheduler \cite{loshchilov_sgdr_2017}.
As mentioned in \refSec{Sampling}, we initialize the sampler with a  low-discrepancy sequence and the guess-reuse techniques proposed in \cite{nielsen2015optimal}. More specifically, we start from training $\sampleCount_0=1$ samples with multiple guesses for the best loss in train set. Then, when we train more $\sampleCount_i = 2\sampleCount_{i-1}$ samples, we reuse those $\sampleCount_{i-1}$ learned samples and initialize the other half with a 3D Sobol sequence.  

In practice, the two for-loops implementing the meta outer loop in \refAlg{Method} could operate batched to improve parallelism and to smooth gradients.
We experimented with meta-batching these loops, but observed no benefit, and hence set the meta-batchsize to one. So this layer of complexity is omitted from the pseudo-code for clarity.

\change{Moreover, we have manually adjusted the learning rate for the \methodRandom method of the \modelNeural model. In \cite{sztrajman2021neural}, the authors use $5\times10^{-4}$, whereas we use $1\times10^{-3}$. Without this change, \methodRandom in \refFig{Teaser} would be entirely black.}

\mysection{Evaluation}{Evaluation}

Our evaluation uses one methodology (\refSec{Methodology}) to produce qualitative (\refSec{Qualitative}) and quantitative (\refSec{Quantitative}) results, which are complemented by some final ablation experiments (\refSec{Ablations}).

\mysubsection{Methodology}{Methodology}
Our evaluation is on 
i) a \emph{dataset}, involving
ii) several \emph{metrics} to measure success, 
iii) \emph{methods} to learn a model and 
iv) \emph{models} describing \acp{BRDF}.
We will now detail all of these four aspects.

\myparagraph{Dataset}
We use the popular MERL \cite{matusik2003data} dataset for our experiments. 
MERL consists of 100 measured \acp{BRDF}, where each \ac{BRDF} is composed of $90\times90\times180 = 1,458,000$ angular configurations $(\theta_h, \theta_d, \phi_d)$ in Rusinkiewicz parametrization and one RGB reflectance per triplet. 
The measured \acp{BRDF} range from diffuse to highly specular, and we organize our data in the classic random $80\%-20\%$ train-test split. Moreover, we also test our approach on the additional eight BRDFs provided by \cite{nielsen2015optimal} \change{and the BRDF data from the RGL material database \cite{dupuy2018adaptive}}. 

\myparagraph{Metrics}
We employ four different metrics.
The first is the mean absolute logarithmic error of the cosine weighted BRDF values, which we use as our optimization loss (\textsc{Loss} in \refAlg{Method}) and hence refer to as the metric ``Loss''.
The other three are image-based DSSIM, L2, and PSNR, for which we render the BRDF on a sphere under environment illumination.
For Loss, DSSIM and L2, less is better, whereas for PSNR, higher values are better. 
All values are reported on unseen test BRDFs, \ie neither method has had access to any of the evaluation data during training. 

\myparagraph{Methods}
We compare three training paradigms:
\methodRandom denotes conventional NN training (as in \cite{sztrajman2021neural}), following \refSec{Random}.
\methodMeta follows \cite{fischer2022metappearance} (for details, cf. \refSec{Meta}), and \methodOur is described in \refSec{Sampling}. As we use a closed form solution for the \modelLinear model, there is no counterpart of \methodMeta. We hence employ the condition number optimization proposed in \cite{nielsen2015optimal} as the baseline to compare against, which is referred to as \methodNielsen.
All methods are compared under \textit{equal time}, \ie using the same number of gradient steps, unless said otherwise.
As for \methodRandom and \methodMeta, we report average results on five independent experiments with different random seeds. 

\myparagraph{Models}
We study applications to all \change{four} different models,
\modelPhong, \change{\modelCookTorrance}, \modelLinear (using five basis functions) and \modelNeural as explained in \refSec{Models}.
Note that \modelLinear is not identical to the specific combination of method and model in \cite{nielsen2015optimal}, which we study separately. 

\mysubsection{Qualitative results}{Qualitative}
The key qualitative results are shown in the right part of \refFig{Teaser}, where different methods are applied to learn \modelNeural for different BRDFs for an equal sample count of $\sampleCount = 64$ at equal time. 
At this point, a common learner, \methodRandom, has not made much progress from the init, which, under the uniform initialization proposed in the original publication \cite{sztrajman2021neural}, results in the black BRDF depicted in the first row.
The second row shows \methodMeta, that has learned a more informative init (the pink sphere on the left), and manages to converge to different materials with a low number of samples (compared to \cite{sztrajman2021neural}, \cite{hu2020deepbrdf} or \cite{fischer2022metappearance}).
However, the appearance is a bit ``stereotype'', i.e., the differences between BRDFs mainly happen via color changes, but the specularity does not match, and the nuances in glossiness are not picked up correctly either. 
Looking at the fourth column, \methodOur, we see that the very different appearance characteristics (e.g., highlights, their shape, color, gloss, etc.) from the reference are faithfully reproduced and that \methodOur matches the reference in the last row most closely.

\refFig{MethodsModels_8samples} shows a similar comparison for which we contrast all methods with all models for a diffuse BRDF at a very low sample count of $\sampleCount=8$ samples only (upper part).
The effects we saw previously at $\sampleCount=64$ samples become even more pronounced: for models \modelPhong, \modelCookTorrance and \modelLinear, we see that \methodRandom completely fails to faithfully recover the BRDF, and even advanced learning methods like \methodMeta struggle with this highly constrained setting (top and bottom row).
The lower part in \refFig{MethodsModels_8samples} shows the same configuration with a specular BRDF, for which we utilize $\sampleCount=32$ samples, as specular samples are harder to optimize. Again, \methodOur is closest to the reference. Overall, across all models, \methodOur performs best, which shows that our meta-sampling truely gathers valuable information that helps to accelerate the fitting process over drawing random samples. 

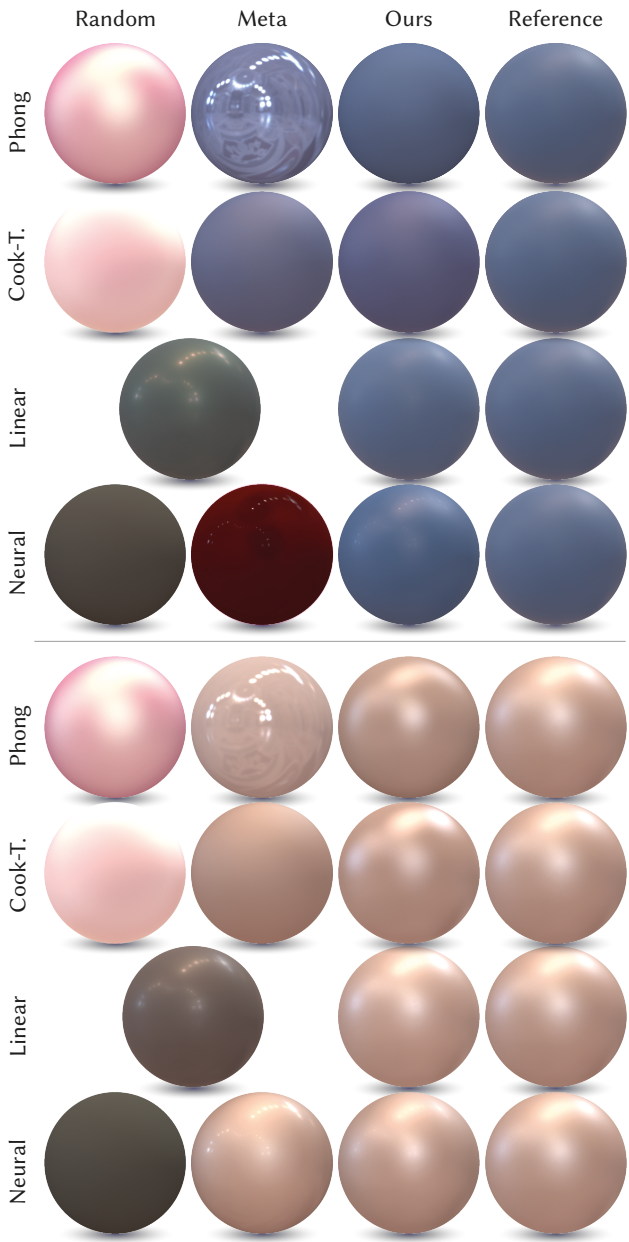
\myfigure{MethodsModels_8samples}{Result for all methods \textbf{(horizontal)} and models \textbf{(vertical)} at equal sample count for the test-set BRDFs \texttt{blue-rubber} (top) and \texttt{silver-paint} (bottom) at $\sampleCount=8$ and $\sampleCount=32$, respectively.}

\newcommand{\metricName}[1]{%
\multicolumn1c{%
\makebox[\widthof{DSSIM}][c]{\tiny #1}
}}

\newcommand{\noData}{\multicolumn1c{---}}
\newcommand{\winner}[1]{\textbf{#1}}
\renewcommand{\arraystretch}{1.1}   
\begin{table*}[]
    \renewcommand{\tabcolsep}{0.017cm}
    \centering
    \caption{Mean test-set performance of all methods on all models at $\sampleCount=8$ samples according to different metrics.}
    \label{tab:SameSampleCount}
    \begin{tabular}{r rrrr rrrr rrrr rrrr}
    \toprule
    \multicolumn1c{Model $\rightarrow$}&
    \multicolumn4c{\modelPhong}&
    \multicolumn4c{\modelCookTorrance}&
    \multicolumn4c{\modelLinear}&
    \multicolumn4c{\modelNeural}\\
    \cmidrule(lr){2-5}
    \cmidrule(lr){6-9}
    \cmidrule(lr){10-13}
    \cmidrule(lr){14-17}
    \multicolumn1c{Method $\downarrow$}
    &
    \metricName{$\downarrow$ Loss}&
    \metricName{$\downarrow$ DSSIM}&
    \metricName{$\downarrow$ L2}&
    \metricName{$\uparrow$ PSNR}&
    
    \metricName{$\downarrow$ Loss}&
    \metricName{$\downarrow$ DSSIM}&
    \metricName{$\downarrow$ L2}&
    \metricName{$\uparrow$ PSNR}&
    
    \metricName{$\downarrow$ Loss}&
    \metricName{$\downarrow$ DSSIM}&
    \metricName{$\downarrow$ L2}&
    \metricName{$\uparrow$ PSNR}&
    
    \metricName{$\downarrow$ Loss}&
    \metricName{$\downarrow$ DSSIM}&
    \metricName{$\downarrow$ L2}&
    \metricName{$\uparrow$ PSNR}
    \\
    \midrule
    \methodRandom&
    0.085&
    0.103&
    0.093&
    12.54&

    \change{0.098}&
    \change{0.123}&
    \change{0.124}&
    \change{10.75}&
    
    \multirow{2}{*}{0.028}&
    \multirow{2}{*}{0.029}&
    \multirow{2}{*}{0.007}&
    \multirow{2}{*}{23.49}&
    0.065&
    0.105&
    0.067&
    13.38
    \\
    \methodMeta&
    0.084&
    0.068&
    0.026&
    18.82&

    \change{0.044}&
    \change{0.039}&
    \change{0.007}&
    \change{24.21}&
    
    &
    &
    &
    &
    0.059&
    0.125&
    0.052&
    15.23
    \\
    \methodNielsen&
    \noData&
    \noData&
    \noData&
    \noData&

    \noData&
    \noData&
    \noData&
    \noData&
    
    0.016&
    0.012&
    0.002&
    28.45&
    \noData&
    \noData&
    \noData&
    \noData
    \\
    \methodOur&
    \winner{0.044}&
    \winner{0.034}&
    \winner{0.006}&
    \winner{24.51}&

    \winner{\change{0.039}}&
    \winner{\change{0.029}}&
    \winner{\change{0.003}}&
    \winner{\change{27.09}}&
    
    \winner{0.011}&
    \winner{0.009}&
    \winner{0.001}&
    \winner{33.70}&
    \winner{0.035}&
    \winner{0.033}&
    \winner{0.005}&
    \winner{27.50}
    \\
    \bottomrule
    \end{tabular}
\end{table*}
\renewcommand{\arraystretch}{1}

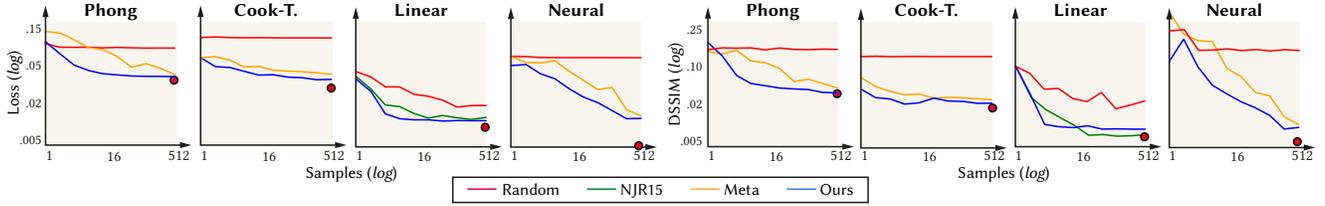
\mycfigure{QualityVsSamples}{Performance (\textbf{vertical}, log scale) of different learning methods (\textbf{colors}) for different models according to different metrics (Log. MAE in BRDF space and image-based DSSIM in every pair, lower is better) depending on the sample count (\textbf{horizontal}, log scale).
The red dot indicates the theoretical optimum, when giving the model \defaultOrders of magnitude more samples, i.e., compute and acquisition time.}

\mysubsection{Quantitative results}{Quantitative}

\myparagraph{Same-sample count}
Our main quantitative result is seen in \refTbl{SameSampleCount}, where we study a same-sample-count ($\sampleCount=8$) setting for all models (\change{four} major column blocks) and all methods (rows) under all metrics (four-blocks of columns).
We see, that for all models according to all metrics, \methodOur is able to improve upon \methodMeta, which again is an improvement over \methodRandom.
The improvement is ranging between a factor of two and four. 
Most importantly, our method is able to improve the  quality of a neural \ac{BRDF} model by co-optimizing the sample pattern, the main contribution of this work.
As \refTbl{SameSampleCount} only displays the average outcome across all test BRDFs, we further display each method's performance on all individual BRDFs in the test set at $\sampleCount=8$ in \refFig{Zipfplots}. 
To this end, we sort the resulting BRDFs according to the loss values achieved by \methodOur, i.e., each horizontal point in \refFig{Zipfplots} is a BRDF, and contrast them against the other methods. 
We see that we do not only achieve better mean performance, as reported in \refTbl{SameSampleCount}, but outperform the other methods for every BRDF. 

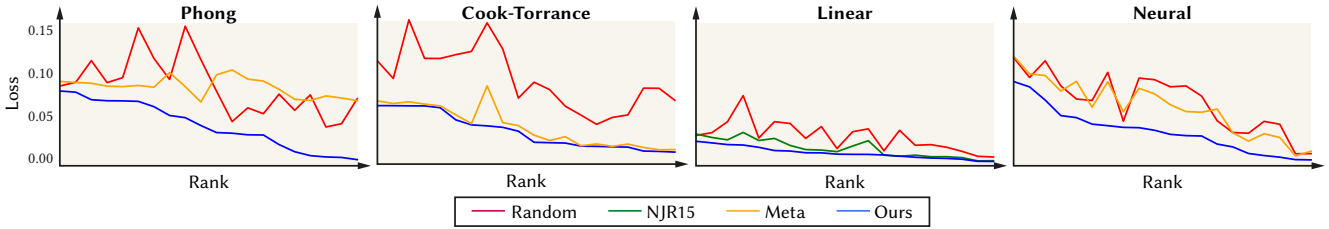
\mycfigure{Zipfplots}{Test set loss (\textbf{vertical}) per BRDF (\textbf{horizontal}), sorted based on the results of \methodOur in decreasing order.}

\myparagraph{Same-quality}
In \refFig{QualityVsSamples} we visualize the influence of sample count on performance, for all models on all methods, according to two representative metrics, Loss in BRDF space and DSSIM in image space.
In each plot, the horizontal axis is the sample count in log scale between 1 and 512.
The vertical axis, also in log scale, is scaled identically per metric, so as to enable comparisons between the individual models. 
We show the upper quality bound for each model for an unlimited number of samples and time as a red dot in the lower right. 
This is the target for all methods economizing on sample count to chase.
Always, less is better.
The red lines show \methodRandom and confirm our qualitative findings from \refSec{Qualitative}, as it does not manage to do much at such low sample counts, even with many steps.
The yellow lines show \methodMeta, which has seen many inner-loop trainings and hence can learn to cope with fewer samples, but has no way to change \textit{what} is sampled.
The blue line, denoting \methodOur, consistently performs better across the entire sample range, documenting that the improvement claimed in the previous table and figures generalizes to all sample counts.
This view is made explicit in \refTbl{SameQuality}, which shows how many more samples are required to achieve the quality of \methodOur for all models and all metrics if the sample count is fixed to $\sampleCount=8$.
It by now is clear that a common learner will take much longer, but we also see a good improvement of more than one order of magnitude in sample effort reduction for \methodMeta.
The improvement for the simple \modelPhong model seems to be larger than for more complex models.

For the \modelLinear model, we also computed the performance of \methodNielsen, who equally assume a linear BRDF model, on our test data (green line).
Essentially, this compares using their method of selecting samples to our method of selecting samples for a comparable model (PCA).
In image space, we see that for \change{larger} sample counts ($\sampleCount \geq 16)$, their way of selecting samples is slightly better than \methodOur, whereas for sample counts below 16, our method is superior. 
Interestingly, this finding does not transfer to BRDF space, where we consistently outperform \methodNielsen for all \sampleCount.
We hence assume that their optimization method (minimizing the condition number) is more closely related to human perception than our training loss, and hence achieves a lower DSSIM.
Also, \methodNielsen is not yet as good as the full \modelNeural, to which our method extends.
This implies that the benefit of our method depends on the target model: When the aim is to get most of the (many) \ac{BRDF} samples regardless of model complexity and usefulness (PCA is not compact, not fast to evaluate, etc), \methodNielsen can be superior.
If a compact and efficient model, \eg \modelNeural, is to be used with only few samples, \methodOur is to be preferred.
We have repeated similar experiments with 240 instead of five basis functions, leading to similar results, but with the additional disadvantages in storage, train and test compute requirements.

\begin{wraptable}{r}{5.2cm}
    \renewcommand{\tabcolsep}{0.02cm}
    \centering
    \caption{Comparison of \methodOur and \protect\cite{nielsen2015optimal} on their dataset at $\sampleCount=2$.}
    \label{tab:Nielsen}
    \begin{tabular}{r rrrr}
    \toprule
    &
    \metricName{$\downarrow$ Loss}&
    \metricName{$\downarrow$ DSSIM}&
    \metricName{$\downarrow$ L2}&
    \metricName{$\uparrow$ PSNR}\\
    \midrule
    \methodNielsen&
    0.023&
    0.128&
    0.003&
    26.28\\
    \methodOur&
    \winner{0.018}&
    \winner{0.009}&
    \winner{0.002}&
    \winner{30.02}\\
    \bottomrule
    \end{tabular}
\end{wraptable}%

\myparagraph{Additional BRDF data}
We also compared to \methodNielsen on their dataset of 8 additional BRDFs in \refTbl{Nielsen} and \refFig{Results_NielsenDataset}.
At similar sample counts, our method performs better according to all metrics.
\change{We see that our method, albeit trained on MERL only, can find sample patterns that generalize reliably to BRDFs from a dataset unseen during training, even if these were acquired with a different acquisition setup.}

\myfigure{Results_NielsenDataset}{Results on data from \cite{nielsen2015optimal} at $\sampleCount=2$ samples.}

\pagebreak
\change{
We also compare to the isotropic part of RGL material database \cite{dupuy2018adaptive} and show the results in \refTbl{RGLeval}. Although the overall loss slightly increases due to data shift, our meta-learned samples over the MERL corpus generalize to the new dataset and lead to better reconstructions.}

\begin{table}[htb]
    \renewcommand{\tabcolsep}{0.2cm}
    \centering
   \caption{Evaluation of \methodOur on the RGL database \protect\cite{dupuy2018adaptive} at $\sampleCount=8$.}
    \label{tab:RGLeval}
    \begin{tabular}{rr rrrr}
    \toprule
    &
    &
    \metricName{$\downarrow$ Loss}&
    \metricName{$\downarrow$ DSSIM}&
    \metricName{$\downarrow$ L2}&
    \metricName{$\uparrow$ PSNR}\\
    \midrule
    \multirow{2}{*}{\modelPhong} &
    \methodMeta &
    \change{0.095}&
    \change{0.070}&
    \change{0.025}&
    \change{17.980}\\
    &
    \methodOur&
    \change{\winner{0.044}}&
    \change{\winner{0.040}}&
    \change{\winner{0.009}}&
    \change{\winner{22.686}}\\
    \midrule
    \multirow{2}{*}{\modelCookTorranceShort} &
    \methodMeta &
    \change{0.068}&
    \change{0.059}&
    \change{0.026}&
    \change{17.824}\\
    &
    \methodOur&
    \change{\winner{0.067}}&
    \change{\winner{0.051}}&
    \change{\winner{0.009}}&
    \change{\winner{21.563}}\\
    \midrule
    \multirow{2}{*}{\modelLinear} &
    \methodNielsen &
    \change{0.050}&
    \change{0.068}&
    \change{0.025}&
    \change{21.897}\\
    &
    \methodOur&
    \change{\winner{0.030}}&
    \change{\winner{0.041}}&
    \change{\winner{0.016}}&
    \change{\winner{23.849}}\\
    \midrule
    \multirow{2}{*}{\modelNeural} &
    \methodMeta &
    \change{0.074}&
    \change{0.168}&
    \change{0.110}&
    \change{12.228}\\
    &
    \methodOur&
    \change{\winner{0.044}}&
    \change{\winner{0.062}}&
    \change{\winner{0.011}}&
    \change{\winner{21.889}}\\
    \bottomrule
    \end{tabular}
\end{table}

\renewcommand{\arraystretch}{1.1}   
\begin{table*}[]
    \renewcommand{\tabcolsep}{0.02cm}
    \centering
    \caption{Relative number of samples required by other methods to achieve our quality at $n=8$ samples. Here, less is better for all metrics.}
    \label{tab:SameQuality}
    \begin{tabular}{r rrrr rrrr rrrr rrrr}
    \toprule
    \multicolumn1c{Model $\rightarrow$}&
    \multicolumn4c{\modelPhong}&
    \multicolumn4c{\modelCookTorrance}&
    \multicolumn4c{\modelLinear}&
    \multicolumn4c{\modelNeural}
    \\
    \cmidrule(lr){2-5}
    \cmidrule(lr){6-9}
    \cmidrule(lr){10-13}
    \cmidrule(lr){14-17}
    \multicolumn1c{Method $\downarrow$}&
    \metricName{Loss}&
    \metricName{DSSIM}&
    \metricName{L2}&
    \metricName{PSNR}&
    \metricName{Loss}&
    \metricName{DSSIM}&
    \metricName{L2}&
    \metricName{PSNR}&
    \metricName{Loss}&
    \metricName{DSSIM}&
    \metricName{L2}&
    \metricName{PSNR}&
    \metricName{Loss}&
    \metricName{DSSIM}&
    \metricName{L2}&
    \metricName{PSNR}
    \\
    \midrule
    \methodRandom&
    $10^5\times$&
    $10^5\times$&
    $10^5\times$&
    $10^5\times$&

    \change{$10^5\times$}&
    \change{$10^5\times$}&
    \change{$10^5\times$}&
    \change{$10^5\times$}&
    
    \multirow{2}{*}{$10^5\times$}&
    \multirow{2}{*}{$10^5\times$}&
    \multirow{2}{*}{$10^5\times$}&
    \multirow{2}{*}{$10^5\times$}&
    $10^5\times$&
    $10^5\times$&
    $10^5\times$&
    $10^5\times$
    \\
    \methodMeta&
    48$\times$&
    37.5$\times$&
    24$\times$&
    8$\times$&

    \change{$8\times$}&
    \change{$64\times$}&
    \change{$64\times$}&
    \change{$64\times$}&
    
    &
    &
    &
    &
    4$\times$&
    6$\times$&
    8$\times$&
    6.5$\times$
    \\
    \methodNielsen&
    \noData&
    \noData&
    \noData&
    \noData&

    \noData&
    \noData&
    \noData&
    \noData&
    
    8$\times$&
    3$\times$&
    3$\times$&
    4$\times$&
    \noData&
    \noData&
    \noData&
    \noData
    \\ 
    \methodOur&
    \winner{1$\times$}&
    \winner{1$\times$}&
    \winner{1$\times$}&
    \winner{1$\times$}&

    \change{\winner{1$\times$}}&
    \change{\winner{1$\times$}}&
    \change{\winner{1$\times$}}&
    \change{\winner{1$\times$}}&
    
    \winner{1$\times$}&
    \winner{1$\times$}&
    \winner{1$\times$}&
    \winner{1$\times$}&
    
    \winner{1$\times$}&
    \winner{1$\times$}&
    \winner{1$\times$}&
    \winner{1$\times$}
    \\
      
    \bottomrule
    \end{tabular}
\end{table*}
\renewcommand{\arraystretch}{1}   

\myfigure{ResultsSubset}{
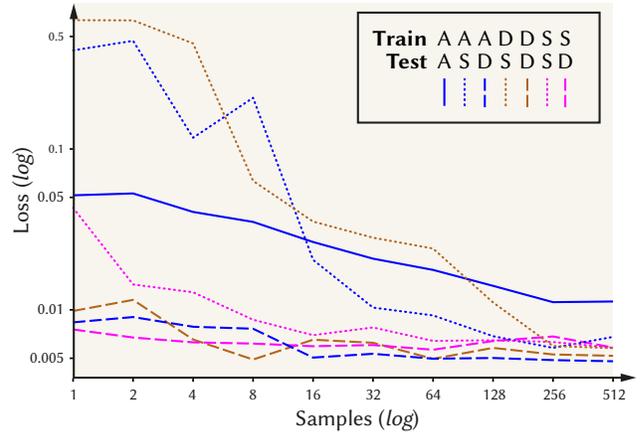
Error (vertical, less is better) at different numbers of samples (horizontal) for different variants of train and test subsets (colors, line styles).
While the training subset corresponds to colors, the test subset maps to line styles (see legend).
For a discussion, please see ``Subsets'' in \refSec{Ablations}.}

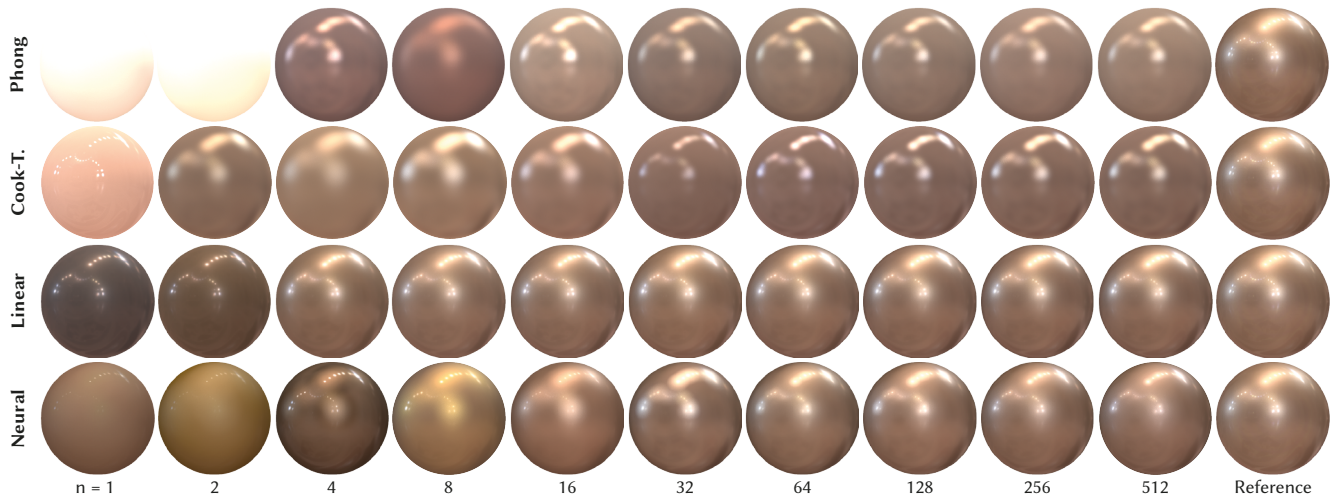
\mycfigure{IncreasingSamples}{We show the results of using our proposed meta-sampling for the training of each model (vertical) on the BRDF \texttt{two-layer-gold} for an increasing number of samples (horizontal). Note that each result is an individual training, not a progression of one training along a row.}

\myparagraph{Sample count}
Here we explore the performance of our sampler \samples with only a limited budget of samples. 
We visualize this for all models in \refFig{IncreasingSamples}, where each column depicts the outcome of a full training run with the number of samples annotated below. 
In general, all methods increase their final quality with more samples. 
We specifically picked a specular BRDF, as these are harder to optimize for (for diffuse, all samples could be in a similar spot). 
At roughly $\sampleCount=8$ samples, we can see first highlights developing. 
Moreover, once a certain quality is reached, \eg, $\sampleCount \geq 64$, the increase in performance with more samples starts diminishing.
This is confirmed by \refFig{QualityVsSamples}, especially for \modelLinear and \modelPhong, and further validates our approach and theory, as in expectation, i.e., $\sampleCount \rightarrow \infty$, the optimization of samples does not matter. 

\myparagraph{Convergence}
We further investigate whether meta-sampling will converge to a solution no worse than pure random sampling.
Due to memory constraints, we cannot directly train \methodOur with the same number of samples as \methodRandom, like 10,000, as that would require back-propagation to 10,000 nested samplings and network executions.
Fortunately, a method much simpler is already better than \methodRandom: if we first run \methodMeta or \methodOur for 20 steps with 32 learned samples, followed by 10,000 steps with 512 random samples, \methodOur already performs better than 20+10,000 steps with 512 purely random samples (0.0117 Loss for \methodRandom, 0.0067 for \methodMeta and 0.0065 for \methodOur).
This confirms that \methodOur does not hamper convergence, but instead, improves it.

\pagebreak
\change{Moreover, when repeating the setting of \refTbl{SameSampleCount} with randomized Quasi-Monte-Carlo initializations (five independent meta-sampling trainings), the standard-deviation for \methodOur is less than 5\% of the total error reported in \refTbl{SameSampleCount}. 
We thus conclude that our method consistently converges to similarly good samples. 
Moreover, when iterating the sampler- and model-training once more after convergence, the Loss of \methodOur improves by 0.9/2.86\% over the values reported in \refTbl{SameSampleCount} for \modelPhong and \modelNeural, respectively.}

\myparagraph{Actual patterns}
The result of meta-sampling for all three methods can be seen in \refFig{Pattern} for 8 samples, and in \refFig{Teaser} for 64 samples and the neural model.
We found it difficult to assign interpretation to those patterns, other than that they seem to group samples and seem not to have the tendency to fill space evenly.

\mymath{\numrandom}{r}
\change{
We verify our learned pattern's robustness by randomly changing points and evaluating the resulting variance. 
We report the increase in loss with respect to \refTbl{SameSampleCount} in \refTbl{robustness} for \numrandom random out of 8 learned samples, averaged over five independent runs.
For all configurations, the loss increases, especially notable for the \modelLinear model, indicating the specific sample positions are relevant.}

\begin{table}[htb]
    \renewcommand{\tabcolsep}{0.3cm}
    \centering
    \caption{Loss-increase for \numrandom random out of eight learned samples.}
    \label{tab:robustness}
    \begin{tabular}{r rrrr}
    \toprule
    &
    \modelPhong&
    \modelCookTorranceShort&
    \modelLinear&
    \modelNeural\\
    \midrule
    $\numrandom=2$ &
    \change{8.3\%}&
    \change{6.9\%}&
    \change{13.9\%}&
    \change{7.3\%}\\
    $\numrandom=4$ &
    \change{12.6\%}&
    \change{10.5\%}&
    \change{21.4\%}&
    \change{14.2\%}\\
    \bottomrule
    \end{tabular}
\end{table}

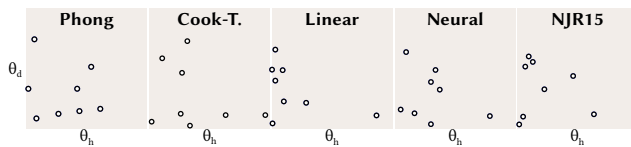
\myfigure{Pattern}{Projections of our meta-sampling patterns at $\sampleCount=8$ for all models. For reference, we also show the pattern from \cite{nielsen2015optimal}.}

\pagebreak
\mysubsection{Ablations}{Ablations}
Here we confirm the relation of our approach to several variants.

\myparagraph{Mean-BRDF importance sampling}
A straightforward approach to reduce the variance of the stochastic gradient is to importance sample for the average BRDF.
To test this, we log-averaged all BRDFs in MERL and then created an inverse \ac{CDF} to use for importance sampling, again employing low-discrepancy samples.
Such a model performs on average more than 4 times worse across all sample counts, in particular for many samples, where it can be over 10 times worse.

\newcommand{\MERLSubset}[1]{$\mathcal{#1}$}
\myparagraph{Subsets}
We now explore how our meta-sampling behaves on selected subsets of the dataset that share specific semantics of MERL.
To this end, we meta-train the model \modelNeural on the full MERL training set as before, but now only use a specific subset for the training of the sampler \samples.
We use two subsets: one with diffuse (\MERLSubset D) and one with specular (\MERLSubset S) materials. 
Moreover, let \MERLSubset A denote the set of all BRDFs.
In the following, we will write $X \times Y$ to denote training of the sampler on subset $X$ and evaluation of its performance on subset $Y$.
In this notation, \MERLSubset{A\times A} is what we already considered in the previous sections: train the sampler on the training set, evaluate its performance on the test set.

We show the results for different train- and test permutations in \refFig{ResultsSubset}.
\MERLSubset{A\times S} performs worse and \MERLSubset{A\times D} performs better than \MERLSubset{A\times A}, indicating specular is harder than diffuse when trained on both.
Importantly, \MERLSubset{S\times S} and \MERLSubset{D\times D}, so methods that were tested on what they were trained on, perform better than \MERLSubset{D\times S} and \MERLSubset{S\times D}, methods that test on something they were not trained for.
In general, that is not an impressive feature for a learned method, but at the same time it proves that the meta-optimized sampling pattern adapts to the characteristics of the dataset, and does not just create some generic useful sample pattern, akin to some perturbed low-discrepancy sequence.


\mysection{Discussion}{Discussion}

\myparagraph{Sample model}
Normalizing Flows \cite{rezende2015variational} look like a well-suited alternative parametrization for our sample model \samples: they generate distributions, produce probability density for samples in the inverse direction and could provide an infinite stream of samples instead of a finite set.
In practice, we have found these properties not relevant, or not applicable to our case, as the generative nature and additional complexity adds further variance to a process that has already two meta-levels and stacks of optimizations.
For tasks other than BRDF, this might become relevant in future work.

\myparagraph{Bias}
Moreover, getting the probability of a sample is important in tasks where we want to retain unbiased estimates, such as in Monte Carlo rendering.
Note that while we sample unevenly, we do not attempt to divide by the probability density to produce unbiased estimates of gradients, as it is not clear whether a biased gradient estimate can ultimately not be better than an unbiased one \cite{diolatzis2022active}.
What matters more is that the outer meta-optimizer sees the effect of those gradients and can factor it  into the optimization by changing the initialization or step sizes.

\myparagraph{Real-world acquisition cost}
We assume a na\"ive cost model of a gonioreflectometer that takes samples in isolation and for which the order of samples does not matter.
\change{
An actual device will deviate from this model for several reasons.
The first is, that multiple samples (slices) can be taken at once by capturing entire images.
Our model assumes samples to be taken in isolation.
The second main simplification is, that the cost of taking a sample while already moving along a trajectory is much lower than changing the direction, which requires acceleration.
Or model assumes that every change of sample direction has the same cost.
For a discussion of BRDF cost properties, please see \cite{guarnera2016brdf}.
}

\myparagraph{Progressivity}
The way we meta-optimize implies that the test-time optimization is good once converged after a fixed number of steps. 
A progressive or interruptible version could be optimized so that it delivers optimal result throughout the entire optimization, by adding all intermediate inner loss values to the meta loss.

\myparagraph{Fixed sample count}
We currently use a fixed-size vector of sample directions, while in practice a method with varying sample counts would be more flexible.
In particular, a method with no limit on the sample count, which eventually converges the same way as random sampling would do.
This could be achieved by meta-learning a generative model of samples, \eg using Normalizing Flow.

\myparagraph{Adaptivity}
The sampling pattern is not on-line \cite{vorba2014line} or adaptive \cite{diolatzis2022active}, but  the same for each \ac{BRDF}.
A pattern that adapts to some other condition, or maybe to the outcome of previous samples, would be a relevant avenue of future work.

\myparagraph{Time and space overhead}
\change{
Meta-learning with MAML \cite{finn2017model} requires the computation of the Hessian-vector product, making it very memory intensive. 
We found that this does not apply so much to our scenario, as the most memory intensive operation is the training of the NBRDF network (2.76 GB VRAM), a property of \cite{sztrajman2021neural} and \cite{fischer2022metappearance}. 
The overhead of meta-sampling is negligible in comparison: for \methodOur at $\sampleCount=10/20$ samples, we require an additional $40.9/51.2$ MB of VRAM, respectively, as we are merely optimizing scalar variables.
There is no overhead at inference time.}

\mysection{Conclusion}{Conclusion}
We have described a method to reduce the number of samples required to fit a non-linear \ac{BRDF} model, such as a \ac{NN}.
To this end, we jointly optimize over the model parameters and the sampling parameters.
Our approach reduce the number of samples required by substantial factors while achieving the same quality.

In future work, we would like to apply the meta-sampling to other domains where samples can be freely taken, such as light field compression \cite{sitzmann2021lfns}, radiance caching \cite{mueller2021neural} or path guiding \cite{vorba2014line}.

\myparagraph{Acknowledgments} \\
\change{We would like to appreciate the anonymous reviewers for their constructive comments and suggestions. We also thank the valuable feedback from Vlastimil Havran, Stavros Diolatzis, and Gilles Rainer.}

\bibliographystyle{eg-alpha}
\bibliography{paper.bib}

\end{document}